# Giant phase modulation in a Mach-Zehnder exciton-polariton interferometer

C. Sturm[1,2], D. Tanese[1], H.S. Nguyen[1], H. Flayac[3], E. Gallopin[1], A. Lemaître[1], I. Sagnes[1], D. Solnyshkov[3], A. Amo[1], G. Malpuech[2] and J. Bloch[1,*]

[1]Laboratoire de Photonique et de Nanostructures, LPN/CNRS, Route de Nozay, 91460 Marcoussis, France.

[2]Universität Leipzig, Institut für Experimentelle Physik II, Linnéstr. 5, 04103 Leipzig, Germany.

[3]Institut Pascal, PHOTON-N2, Clermont Université, Université Blaise Pascal, CNRS, 24 avenue des Landais, 63177 Aubière Cedex, France.

*Correspondence to: jacqueline.bloch@lpn.cnrs.fr.

We report on a new mechanism of giant phase modulation. The phenomenon arises when a dispersed photonic mode (slow light) strongly couples to an excitonic resonance. In such a case, even a small amount of optically injected carriers creates a potential barrier for the propagating exciton-polariton which provokes a considerable phase shift. We evidence this effect by fabricating an exciton-polariton Mach-Zehnder interferometer, modulating the output intensity by constructive or destructive interferences controlled by optical pumping of a micrometric size area. The figure of merit for a π phase shift, defined by the control power times length of the modulated region, is found at least one order of magnitude smaller than slow light photonic crystal waveguides.

The achievement of giant physical phenomena has often been the key ingredient allowing the development of a new technology. For instance, hard disks are based on the discovery of the giant magneto-resistance[1,2], or development of semiconductor laser diodes became possible thanks to the giant reduction of the current threshold enabled by double heterostructures[3,4]. Meanwhile, a giant optical non-linearity which would allow developing of integrated all-optical logical circuits is probably still to be discovered, even though realization of the "optical computer", or, more realistically, of a few integrated all-optical logical elements, was already envisaged in the 70s [5]. Strong progresses have been recently achieved thanks to the use of resonant effects in photonic structures such as photonic crystals. In these micro-systems, a tiny change of the refractive index in a very limited region of space is sufficient to modulate the optical transmission through the device, which allows the realization of all-optical switches with very low switching powers and relatively high repetition rates [6]. The performance of Mach-Zehnder interferometers has been progressively increased using slow light and photonic crystals [7]. However, the physical mechanisms leading to the optical refractive index modulation used in these devices are essentially the same as in the eighties, consisting of carrier induced modulation of the band structure and, in turn, of the optical index, which can be tuned by about $10^{-2}$ with a density of $10^{18}$ carriers per $cm^3$ [8]. Giant amplification of the modulation can be achieved using polariton modes, resulting from the coherent coupling between a propagating photon and an excitation of the media. This coupling drastically affects the properties of the optical mode. For example, the coupling of photons with plasmons allows shrinking the propagating optical mode well below the diffraction limit [9]. The strong coupling of photons with excitons gives rise to exciton-polaritons, which combines the strong non-linearity of an electronic system (due to exciton-exciton interactions) with the large propagation length of photons. Besides the strong coupling, the second key ingredient is the quantization of the photon mode in a microcavity, making it highly dispersive. The first demonstration of this potentiality came with the realization of polaritonic parametric oscillator [10] and amplifier [11,12] where the $\chi^3$ nonlinearity was found orders of magnitude larger than in any other optical systems. Low threshold bistability [13,14] has been demonstrated together with several schemes of optical switches [15,16] and transistors [17,18,19]. These features along with the now well-established polariton condensation in etched microstructures [20], make semiconductor microcavities an excellent platform for integrated photonics [21], where coherent emission, optical guiding and non-linearity can be combined within the same chip.

An interesting possibility of manipulating a polariton flow is to use an external potential created by a non-resonant optical pumping [20,22]. The pump locally creates carriers or excitons which interact directly with polaritons, inducing a potential barrier $V$ which can reach several meVs. This optically induced potential has been used so far to confine polaritons in well-defined regions of space. Another configuration, which has not yet been explored, occurs when the kinetic energy $E_k$ of a propagating polariton flow is larger than $V$. In such a case, the polariton flow is well transmitted but acquires a phase shift difference $\delta\varphi$ as compared to the case without barrier. Considering the simplest case of a square shape potential barrier, the induced phase shift difference is given by:

$$\delta\varphi = L\left(\frac{\sqrt{2mE_k}}{\hbar} - \frac{\sqrt{2m(E_k - V)}}{\hbar}\right) \quad (1)$$

where $m$ is the polariton effective mass, and $L$ the width of the optically induced potential. Taking realistic parameters, such as $E_k = 1.5$ meV, $V = 1$ meV and a polariton mass of $4 \cdot 10^{-5} m_0$, a π phase shift can be realized over a region of 6 μm only. This large value can be obtained because the polariton kinetic energy is comparable with the height of the potential barrier. In other words, thanks to the quantization of the cavity mode, we are dealing with highly dispersed "slow light" for which the impact of a weak barrier, amplified by strong coupling, is maximal.

In this report, we experimentally evidence this giant phase shift effect and illustrate its potentiality in the implementation of a low power integrated photonic circuits. In order to do so, we have fabricated two polariton circuits by etching a high quality GaAs/AlGaAs microcavity [24]: a tennis-racket shaped circuit and a Mach-Zehnder interferometer (MZI). Figure 1A shows a Scanning Electron Microscopy (SEM) image of the "tennis racket" microcavity. The handle of the racket is excited by a non-resonant *cw* laser focused on a 2 μm diameter area with a pumping power above the threshold for polariton condensation. This pumping scheme allows creating a mono-kinetic flow of coherent polaritons, having a kinetic energy of the order of 2.5 meV [20]. This flow splits in two propagating beams when entering the racket which interfere in the opposite part of the racket. The coherence of the condensate is maintained during propagation and interference fringes can be directly observed when imaging the polariton emission (Figure 1C). A control laser beam is focused on the upper part of the racket with a spot size around 7 μm. This laser is strongly blue detuned with respect to the polariton resonances (non-resonant excitation) and locally injects a reservoir of uncondensed excitons. Because of repulsive polariton-exciton interactions, this reservoir induces a potential barrier with a spatial profile reproducing the shape of the control beam and a height proportional to the exciton density [20]. We can deduce the height of the barrier for different values of the control beam power $P_c$, from the measured local blueshift of polariton resonances (inset in Figure 1B). In our experiments, this potential barrier is kept smaller than the kinetic energy of the flow. Thus polaritons can pass through the barrier, but acquire a phase shift as compared to the case without this barrier. Figure 1, D-F, present the interference patterns obtained for increasing $P_c$. The induced phase shift is highlighted by a progressive displacement of the interference fringes and reach values as large as $3\pi$ (see Figure 1F).

This strong non-linearity can be used to control the output in a small sized polariton MZI. To demonstrate the operating regimes of the MZI, we use resonant optical excitation for injection of the incoming polariton flow: this allows control of its energy and polarization. As shown in the following, depending on the energy of the injected polariton flow, different operating regimes can be achieved. To understand these regimes, we have to consider that polariton lateral confinement in a one-dimensional cavity gives rise to the formation of several 1D polariton sub-bands (Figure 2B). The lowest energy sub-band is TM polarized (polarization along the wire), while the second one is TE polarized (polarization perpendicular to the wire).

We first consider resonant injection of TM polaritons with energy below the TE sub-band (cf. Figure 2B). Measured transmission in the outgoing arm versus $P_c$ is shown on Figure 2C. Interference effects induced by the control laser beam are illustrated in the real space images of the polariton emission shown in Figure 2, D-F. For $P_c = 0$, the signals propagating in the two arms interfere constructively at the device output, allowing maximum transmission of the device (Figure 2D). Switching on the control laser, a $\pi$ phase shift is realized for $P_c = 5$ mW, which leads to destructive interference at the output. Pronounced transmission decrease by one order of magnitude is then achieved. Maximum transmission is recovered for $P_c = 10$ mW, corresponding to a $2\pi$ phase shift induced by the control beam. The observed behavior of our device is reproduced using a simulation based on a 2D spinor Schrödinger equation (Figure 2, G-I). We consider the propagation of a coherent wave in a guiding potential, which defines the shape of the circuit, and introducing a barrier potential of various height which reflects the effect of the control laser beam. Simulated images fully reproduce all the interference patterns observed experimentally. This experiment demonstrates that taking advantage of the giant non-linearity, we can access the full range of transmission modulation values for the polariton MZI imposing the phase modulation over a region as small as 7 μm.

Beside the control of the transmission, the present polariton MZI can be used for the realization of a polarization switch. To evidence this effect, we inject TM polaritons with energy above the bottom of the TE sub-band (inset of Figure 3A). Because of the non-adiabaticity of the propagation at the entrance of the interferometer, a part of the TM incoming polaritons is converted to TE-polarized polaritons. This conversion can be understood as originating from the precession of the polariton pseudospin around the effective magnetic fields induced by the photon mode TE-TM splitting [23]. Polaritons entering the upper and lower arms experience opposite effective fields leading to the rotation of the polariton pseudo-spin in opposite directions. As a result, the phases of the TE wave injected in the two arms are opposite, so that the TE-polarized polaritons interfere destructively at the output as clearly visible on Figure 3A. In contrast, the phase of the TM-polarized polaritons is the same in the two arms so that constructive interference occurs at the output. Finally only a TM-polarized wave is transmitted through the device. When the control beam is applied, different phase shifts are induced for TE and TM polarizations. Thus a switching of the polarization of the transmitted signal can be observed when a π phase shift is obtained for TM polarization (for $P_c$ = 7 mW as seen in Figure 3D). Interference between TM waves becomes destructive, whereas interference between TE waves is constructive. For a 2π phase shift, the TM-polarization of the original signal is recovered. Figure 3 summarizes how the linear polarization ratio changes sign when increasing $P_c$. Simulation using the model described above reproduces this polarization control.

We can also demonstrate switching between transverse modes when pumping the system above the third polariton band [24]. More complex schemes can be easily envisaged based on the present mechanism of polariton phase modulation [25]. In the supplementary we simulate a router based on a polariton MZI with two output leads.

To evaluate the performance of our all optically controlled device, we define its figure of merit, as the required control power times length of the modulated region, to obtain a π phase shift. We obtain typically $10^{-2}$ W·μm in our polariton MZI, as compared to values several orders of magnitude larger when the bulk optical non-linearity is used in a waveguide [26], and still ten times better than photonic crystal devices based on slow light [7]. We estimate that the carrier (exciton) density required to drive our polariton MZI is of the order of $10^{10}$ cm$^{-2}$ per quantum wells, and corresponds to a volume density of only $10^{16}$ cm$^{-3}$.

Finally we want to address real potential of this giant phase modulation for applications. First, to reduce radiative losses along the propagation, guided surface polariton modes could be envisaged. These modes should be strongly dispersed (slow light effect) and strongly coupled. Such configuration could be realized in photonic crystal slabs with embedded quantum wells [27] or using laterally confined Bloch surface polaritons [28]. The operating temperature could be raised up to room temperature using large band gap semiconductors, such as GaN or ZnO [29,30] where material quality has strongly improved lately. Regarding the switching time, we expect it to be comparable with the one of the carrier-induced index modulation. So it could be of the order of one hundred to a few ps [16].

In conclusion, we are proposing a new platform for optical signal processing, where coherent emission, optical guiding and non-linearity is obtained within the same material. We demonstrate giant phase modulation of a propagating photonic mode, based on the simultaneous achievement of the slow light effect and strong exciton-photon coupling. This first proof of principle opens the way toward more complex devices such as optical gates based on interferometry [31].

**Acknowledgments:**

The authors are particularly grateful for fruitful discussions with Béatrice Dagens about the MZI design and to Luc Le Gratiet for the SEM measurements. This work was partly supported by the FP7 ITN "Clermont4" (235114) and "Spin-Optronics"(237252), by the ANR project Quandyde (ANR-11-BS10-001), by the french RENATECH network and the RTRA Triangle de la Physique contract "Boseflow1D" and "2012-039T-InterPol". This work was partially supported by a travel grant of the POLATOM Research Networking Program of the European Science Foundation (ESF).

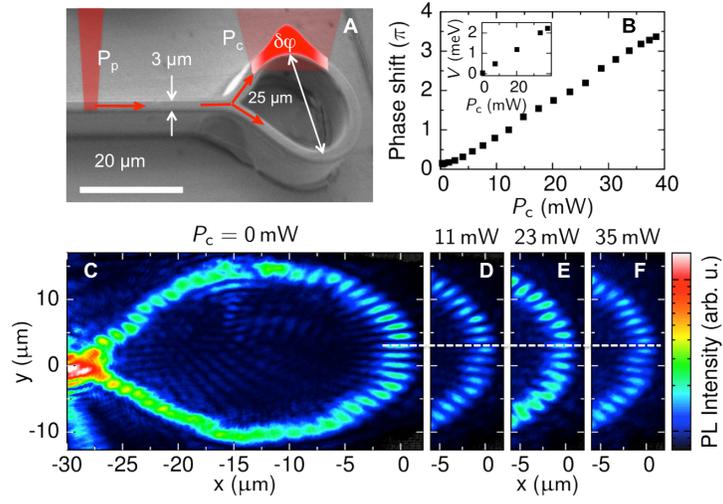

**Fig. 1**. **(A)** SEM image of the tennis racket structure. **(B)** Measured phase shift as a function of $P_c$; inset: measured potential height $V$ as a function of $P_c$. **(C–F)** Real space imaging of polariton emission in the tennis racket device for several values of $P_c$.

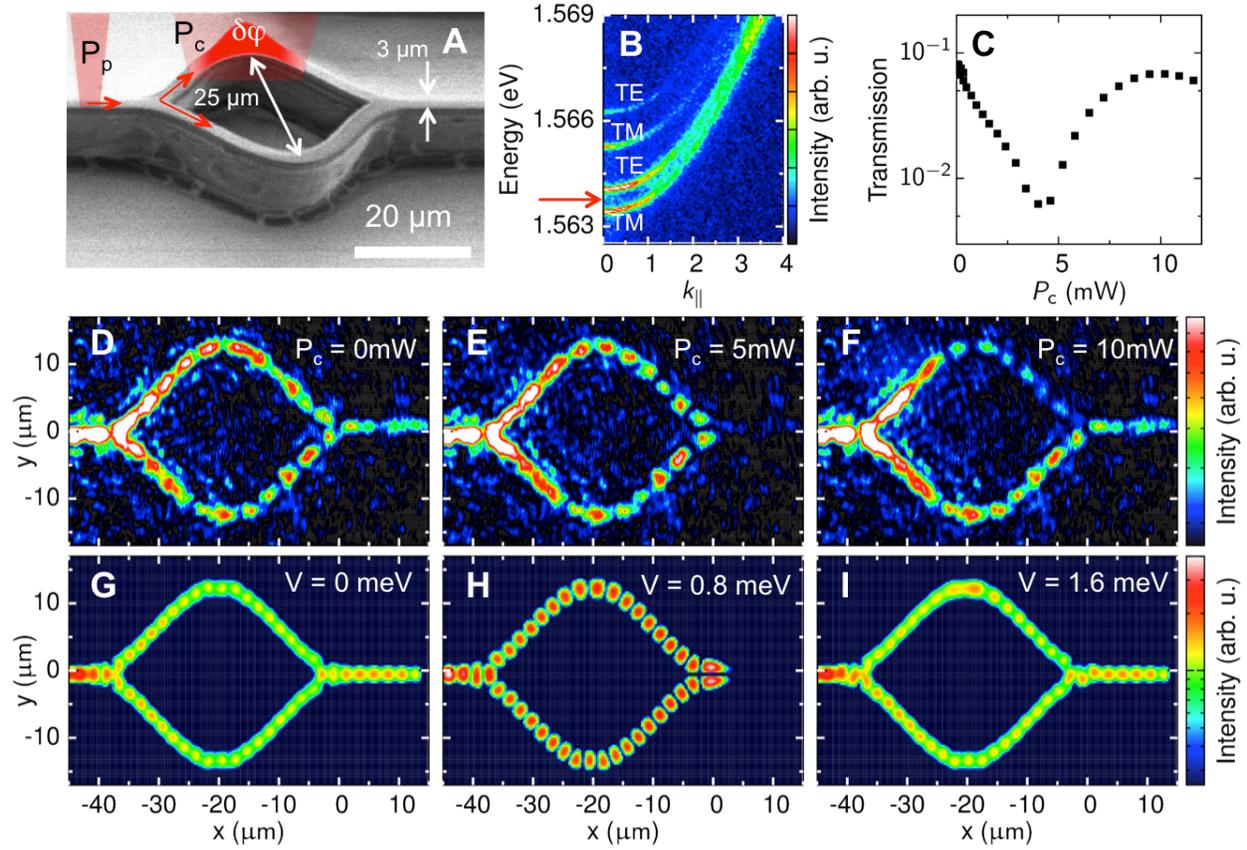

**Fig. 2.** Modulation of the MZI transmission (**A**) SEM image of the polariton MZI. (**B**) Measured polariton dispersion. The red arrow indicates energy of injected polariton flow; (**C**) Measured transmission as a function of $P_c$; (**D – F**) Spatially resolved polariton emission measured for different values of $P_c$; (**G – I**) Calculated emission pattern for different heights of a gaussian ($\sigma$ = 4 µm) induced potential (Parameters: $E_k$ = 1.85 meV, $m = 4 \cdot 10^{-5} m_0$).

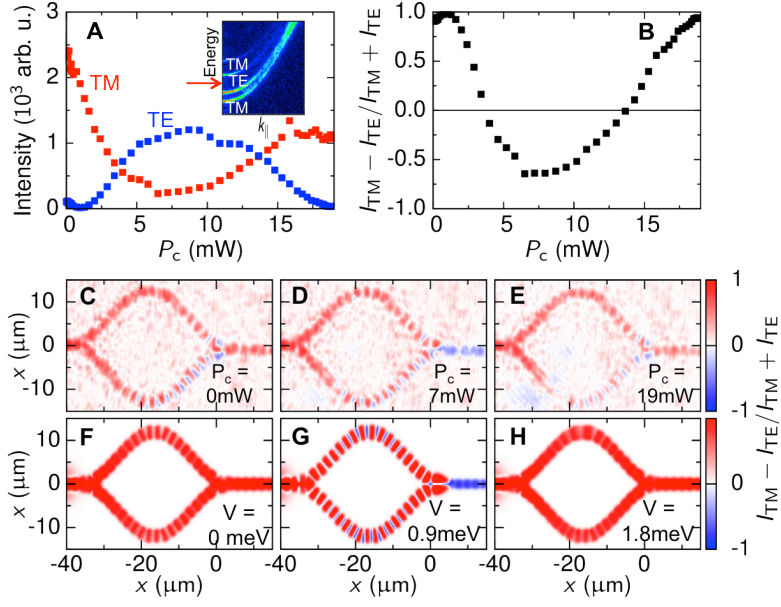

**Fig. 3.** Polarization switch **(A)** Mesured transmitted intensity in (blue) TM and (red) TE polarization as a function of $P_c$; inset: polariton dispersion with a red arrow indicating energy of injected polariton flow; **(B)** Linear degree of polarization of the output beam as a function of $P_c$; **(C – E)** Spatially resolved measured degree of polarization for three values of $P_c$; **(F – H)** Calculated degree of polarization for a splitting between TE-TM bands $E_{k=0}^{TE} - E_{k=0}^{TM} = 0.4\,\text{meV}$, $E_k = 1.85$ meV, $m = 4 \cdot 10^{-5} m_0$ and different heights of a gaussian induced potential ($\sigma = 4\,\mu\text{m}$).

**Supplementary Materials:**

**Sample and measurements**:

The sample consists of a λ/2 $Ga_{0.05}Al_{0.95}As$ cavity embedded between two $Ga_{0.05}Al_{0.95}As$/$Ga_{0.8}Al_{0.2}As$ Bragg mirrors with respectively 40 (32) pairs in the bottom (top) mirror. The nominal quality factor is around $10^5$. Three sets of four 7 nm GaAs quantum wells are inserted at the antinodes of the electromagnetic field resulting in a 15 meV Rabi splitting. The tennis-racket and Mach-Zehnder interferometers were designed using electron-beam lithography and dry etching. In both structures the diameter of the ring was chosen to be 25 μm and the width of the wires as well as of the ring is 3 μm. To optimize efficiency of injection, a smooth Y-shaped design has been used, where the curvature is the one of a lemniscate of Bernoulli. The exciton-photon detuning defined as $E_C(k=0) - E_X(k=0)$ is equal to -4.5 meV (resp. -8 meV) for the tennis-racket (resp. MZI) structure studied in the paper.

Micro-photoluminescence measurements are performed at 10 K. The excitation is provided by two single mode Ti:Sapphire lasers. The far (respectively near) field spectroscopy is obtained projecting the Fourier plane of the detection microscope objective (respectively the sample surface plane) on the entrance slit of a monochromator, placed prior to a nitrogen cooled CCD camera.

**Simulation using 2D spinor Schrödinger equation**:

We simulate a polariton flow inside a 2D waveguide structure corresponding to the experimental interferometer structure. Along one arm of the interferometer we consider an induced 2D gaussian-shaped potential with σ = 4μm. The height $V$ of the potential barrier acting on polaritons is taken $V = |\chi|^2 n_x 6 a_b^2 E_b$ where $|\chi|^2$ is the exciton fraction of the polartion, $n_x$ is the exciton density in the reservoir, $a_b$ is the exciton Bohr radius, $E_b$ the exciton binding energy.

The splitting between transversal and longitudinal polaritonic modes is reproduced in the model by the introduction of an effective in plane magnetic field. It actson the polariton pseudopsin, generating a Zeeman splitting between polaritonic states of different polarization and a sub-bands-structure analogous to the experimental one.

**Modulation of the group index**

The contribution of the slow light effect to our device can be quantified by define an effective group index for the in-plane propagation as $n_g = c/v_g = \hbar c \, \partial k / \partial E_k$, where c is the speed of light and $v_g$ the polariton group velocity. For a kinetic energy $E_k$ of the order of 2.5 meV (resp. 0.6 meV) one gets $n_g$ = 80 (resp. 160). As explained in the main text, the optically induced potential changes the in-plane k vector and therefore inducing a change in the group index given by:

$$\frac{\Delta n_g}{n_g} = \sqrt{\frac{E_k}{E_k - V}} - 1$$

As an example, the experiment shown in Figure 1 corresponds to a change of the group index as large as $\Delta n_g / n_g = 3$ (Figure S1). This strong slow light effect induced by the polariton non-linearity explains the giant phase modulation achieved in our devices.

**Switching between transverse modes:**

In Figure 3 we have shown that by injecting polaritons in the MZI above the second polariton subband, we can control the output polarization of our device via the mixing of the lowest TE[1] and $TM^1$ modes inside the structure. These modes correspond to the lowest transverse modes of each polarization. We are now interested in going a step further and play with the conversion of different transverse modes. Here we inject polaritons in the MZI with energy above the third polariton sub-band ($TM^2$) (see Figure S2A). Such a band corresponds to second order transverse polariton modes with a transverse distribution showing two antinodes with opposite phase, and one node at the center, whereas the lowest subbands ($TM^1$ and $TE^1$) are first order transverse modes with an antinode in the center. Because of the gaussian profile of the excitation beam, essentially $TM^1$ and $TE^1$ polaritons are injected in the interferometer. These two modes have almost the same k vector $k_1$, and so they undergo almost the same phase shift along propagation. At the entrance of the interferometer, part of these polaritons are transferred into the mode $TM^2$, characterized by much smaller k vector $k^2$. Because this $TM^2$ mode presents two out of phase antinodes, it is easy to show that $TM^2$ polaritons injected in the upper arm are out of phase with respect to $TM^2$ polaritons injected in the lower arm. As a result, they undergo destructive interference at the MZI output and only first order polariton modes are transmitted when $P_c = 0$ (see Figure S2C). When applying the control laser beam, $TM^2$ starts to be transmitted and spatially oscillating intensity is obtained at the output. When a $\pi$ phase shift for first order modes is reached, only second order polaritons are transmitted with the characteristic transverse profile (Figure S2E). Good transmission of first order polariton modes is recovered for an induced phase shift of $2\pi$ on the second order mode (Figure S2 G). This switching between different transverse modes is reproduced by our simulations as can be seen in Figure S2, H–L.

**Proposal for a polariton router**:

More complex schemes can be envisaged based on the reported mechanism for polariton phase modulation. Here we propose a polariton router and consider a polariton MZI such as the one described in the paper but with two output leads. Figure S3 shows the results of numerical simulations showing that the control laser beam can switch transmission from one output lead to the other. Indeed changing the phase shift in one of the MZI arm moves the position of the interference fringes in the ring. In Figure S3A, the upper output lead is aligned with a node of the polariton field whereas the lower one is aligned with an antinode. This explains why transmission essentially occurs from the lower output. Opposite situation is obtained in Figure S3B corresponding to a different value of $P_c$. Thus controlling $P_c$, on can switch the output signal from one lead to the other.

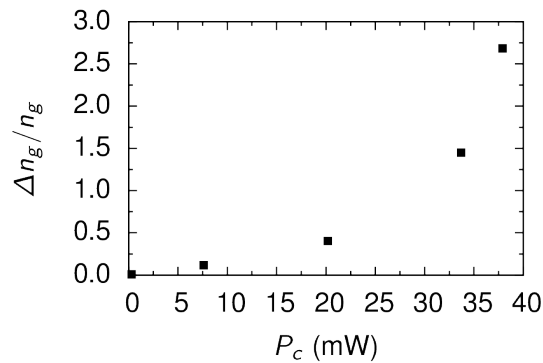

**Figure S1:** Induced change of the group index by the induced potential.

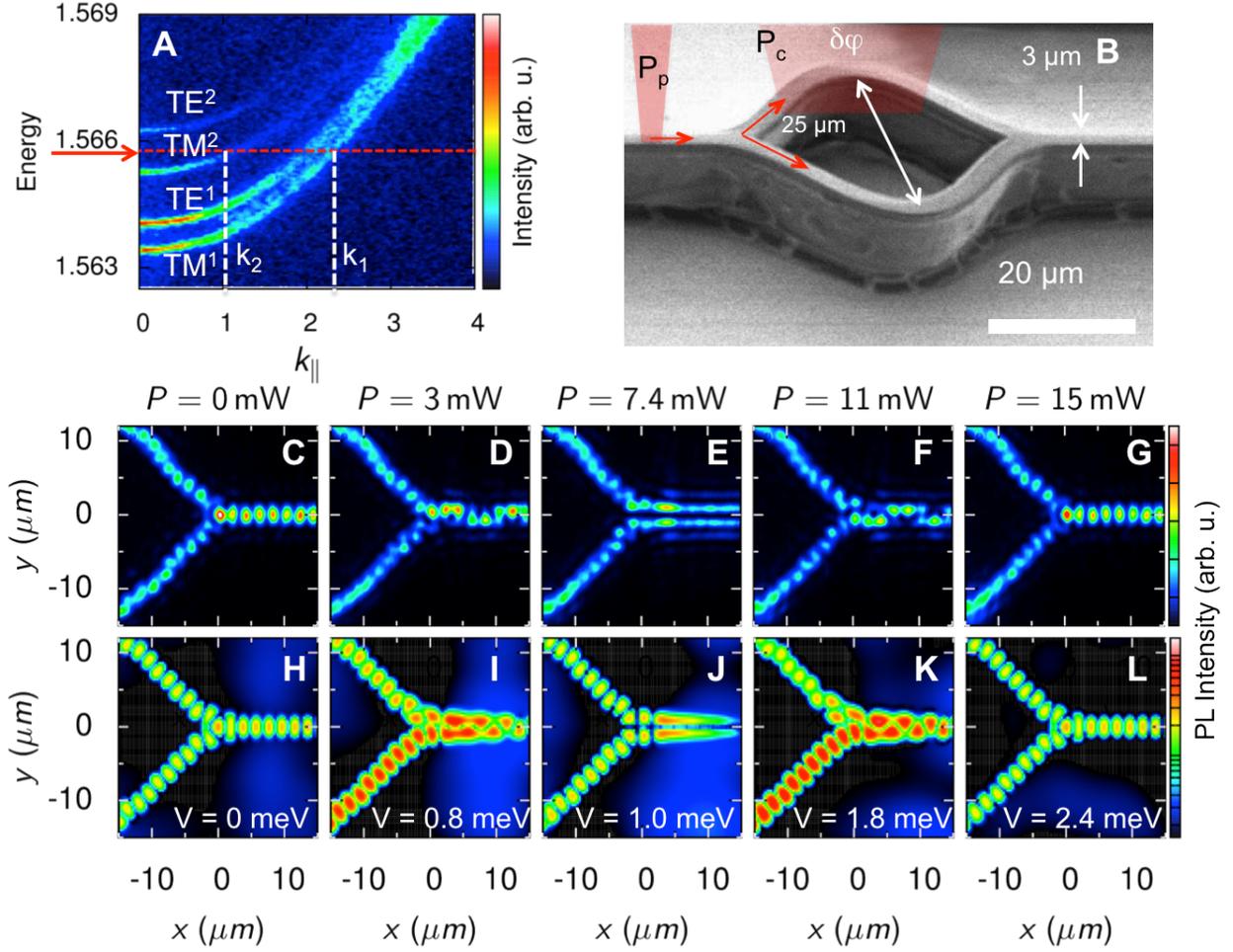

**Figure S2:** Switch between transverse modes **(A)** Measured polariton dispersion showing with red arrow the energy of injected polariton in the present experiment. The white dashed lines indicates the two main wave vectors ($k_1$ and $k_2$) of the first and second order 2D confined polariton modes **(B)** Schematic of the performed experiment. **(C-G)** Spatially resolved emission for five values of $P_c$; **(H-L)** Calculated emission pattern for different heights of the potential barrier (Parameters: $E_k = 3.2\,\text{meV}$, $m = 4\cdot 10^{-5}\,m_0$, $\sigma = 4$ μm).

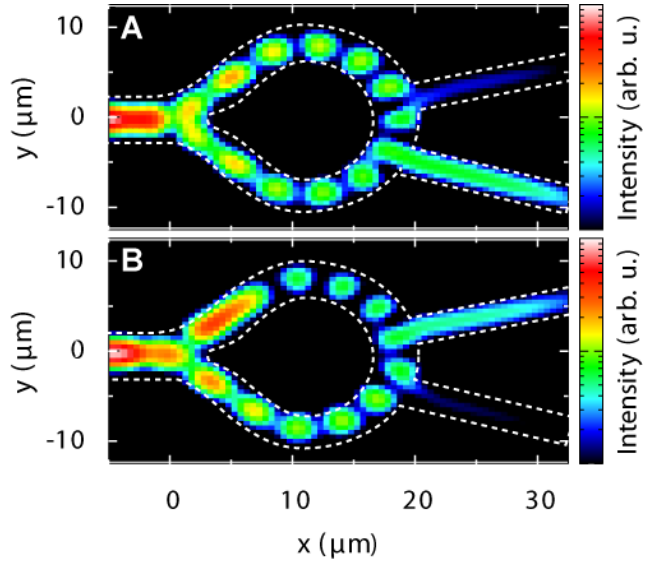

**Figure S3:** Polariton Router (**A and B**) Calculated polariton emission in the polariton router for an induced potential of (A) $V = 0$ meV and (B) $V = 0.8$ meV (Parameters: $E_k = 0.9$ meV, $m = 4 \cdot 10^{-5} m_0$, $\sigma = 4$ μm). The dashed white lines indicate the design of the proposed router.